\documentclass[showpacs,amssymb,twocolumn,aps,nofootinbib]{revtex4}
\usepackage{amsmath}
\usepackage{amstext}
\usepackage{amsopn}
\usepackage{amsfonts}
\usepackage{amssymb}
\usepackage{amssymb}
\usepackage{bbm}
\usepackage{accents}
\usepackage{empheq}
\usepackage{graphicx}
\usepackage{epsf}
\usepackage{graphics}
\usepackage{xcolor}
\usepackage[latin1]{inputenc}

\begin{document}

\title{Operator description for thermal quantum field theories on an arbitrary path in the real time formalism}

\author{Ashok Das$^{a,b}$ and Pushpa Kalauni$^{c}$}
\affiliation{$^a$Department of Physics and Astronomy, University of Rochester,
Rochester, NY 14627-0171, USA}
\affiliation{$^b$ Institute of Physics, Sachivalaya Marg, Bhubaneswar 751005, India}
\affiliation{$^c$Homer L. Dodge Department of Physics and Astronomy, University
of Oklahoma, Norman, OK 73019, USA}

\pacs{11.10.Wx, 11.10.-z, 03.70.+k}
\begin{abstract}
We develop an operator description, much like thermofield dynamics, for quantum field theories on a real time path with an arbitrary parameter $\sigma\,(0\leq\sigma\leq\beta)$. We point out new features which arise when $\sigma\neq \frac{\beta}{2}$ in that the Hilbert space develops a natural, modified inner product different from the standard Dirac inner product. We construct the Bogoliubov transformation which connects the doubled vacuum state at zero temperature to the thermal vacuum in this case. We obtain the thermal Green's function (propagator) for the real massive Klein-Gordon theory as an expectation value in this thermal vacuum (with a modified inner product). The factorization of the thermal Green's function follows from this analysis. We also discuss, in the main text as well as in two appendices, various other interesting features which arise in such a description.
\end{abstract}
\maketitle

\section{Introduction}

There are two commonly used real time formalisms to describe a quantum field theory at finite temperature. The closed time path formalism \cite{schwinger,bakshi,keldysh} uses the path integral method while thermofield dynamics \cite{takahashi, umezawa,umezawa1} has its origin in an operatorial description of thermal quantum field theory. Unlike the imaginary time (Matsubara) formalism \cite{matsubara}, there is a doubling of field degrees of freedom in the real time formalisms \cite{chou,landsman,bellac,das,khanna} which leads to a $2\times 2$ matrix structure for the propagator. Thus, for example, the causal Green's function (the propagator without the factor of $i$) for a real, massive Klein-Gordon field has the momentum space representation in closed time path of the form
\begin{equation}
G^{\rm (CT)} (p) = \begin{pmatrix}
G_{++}(p) & G_{+-} (p)\\
G_{-+} (p) & G_{--} (p)
\end{pmatrix},
\end{equation}
where
\begin{align}
G_{++} (p) & = \frac{1}{p^{2}-m^{2}+i\epsilon} - 2i\pi n_{B} (|p_{0}|) \delta (p^{2}-m^{2}),\notag\\
G_{+-} (p) & = -2i\pi \delta (p^{2}-m^{2})\left(\theta (-p_{0})+ n_{B} (|p_{0}|)\right),\notag\\
G_{-+} (p) & = -2i\pi \delta (p^{2}-m^{2}) \left(\theta (p_{0}) + n_{B}(|p_{0}|)\right),\notag\\
G_{--} (p) & = -\frac{1}{p^{2}-m^{2}-i\epsilon} - 2i\pi n_{B}(|p_{0}|)\delta(p^{2}-m^{2}).\label{CTprop}
\end{align}
Here $n_{B}(|p_{0}|)$ denotes the Bose-Einstein distribution function. (The subscripts $\pm$ refer to the two real branches of the closed time path in the complex plane.) 

In thermofield dynamics, on the other hand, the $2\times 2$ matrix causal Green's function has the momentum space form
\begin{equation}
G^{\rm (TFD)} (p) = \begin{pmatrix}
G_{11}^{\rm (TFD)} (p) & G_{12}^{\rm (TFD)} (p)\\
G_{21}^{\rm (TFD)} (p) & G_{22}^{\rm (TFD)} (p)
\end{pmatrix},
\end{equation}
where the subscripts $1,2$ refer to the two real branches of the time contour. The components have the explicit forms 
\begin{align}
G_{11}^{\rm (TFD)} (p) & = G_{++} (p),\notag\\
G_{12}^{\rm (TFD)} (p) & = -2i\pi e^{\frac{\beta|p_{0}|}{2}} n_{B}(|p_{0}|) \delta(p^{2}-m^{2}) \notag\\
 & = e^{\frac{\beta p_{0}}{2}} G_{+-} (p),\notag\\
G_{21}^{\rm (TFD)} (p) & = -2i\pi e^{\frac{\beta|p_{0}|}{2}} n_{B} (|p_{0}|) \delta(p^{2}-m^{2}) \notag\\
& = e^{-\frac{\beta p_{0}}{2}} G_{-+} (p),\notag\\
G_{22}^{\rm (TFD)} (p) & = G_{--} (p).\label{TFDprop}
\end{align}
Here $\beta$ denotes the inverse temperature in units of the Boltzmann constant. Even though the off-diagonal parts of the Green's functions in \eqref{CTprop} and \eqref{TFDprop} have different forms in the two formalisms, they are known to lead to equivalent results for physical ensemble averages in thermal equilibrium.

\begin{figure}[h]
\includegraphics[scale=0.52]{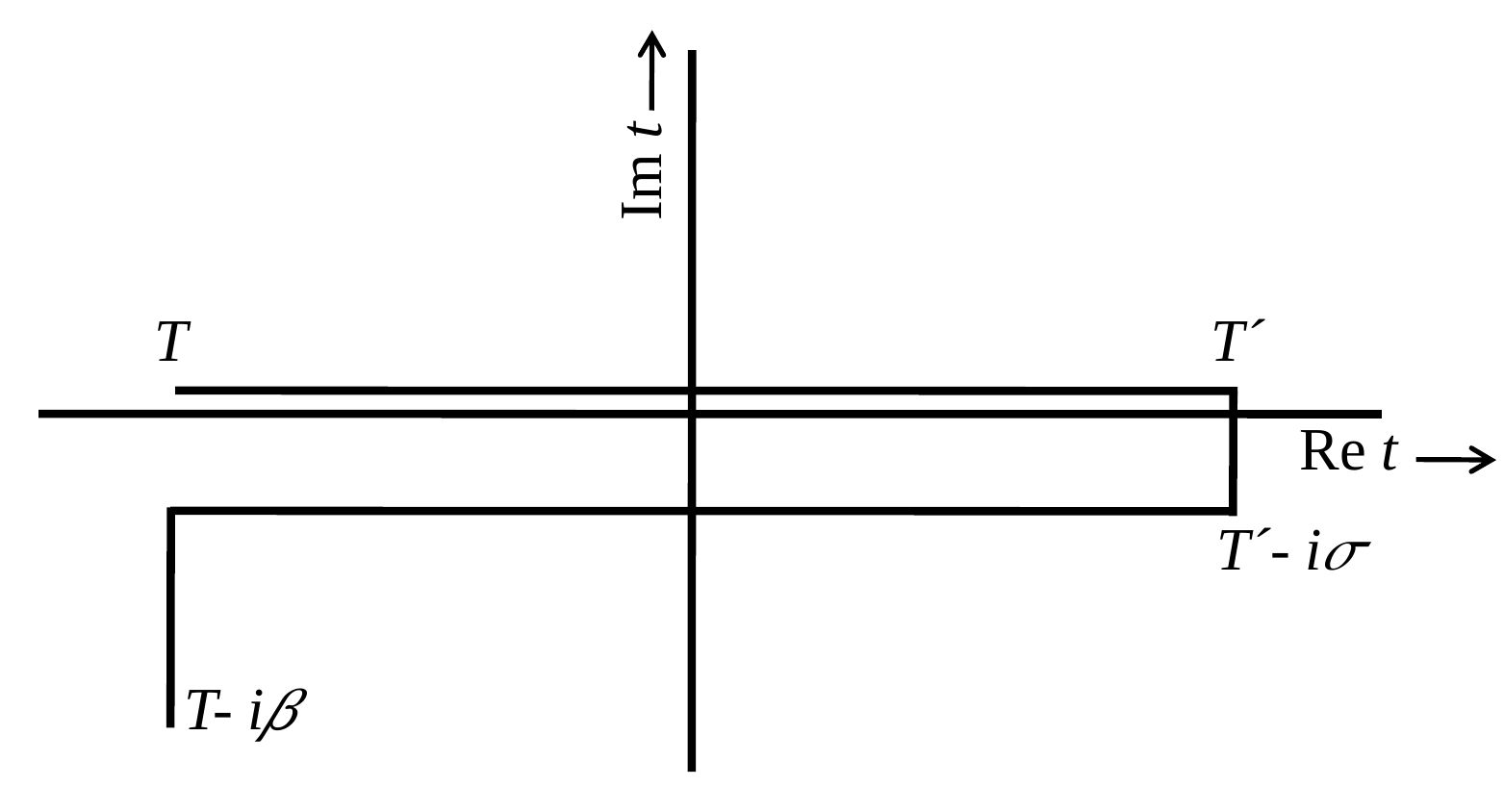}
\caption{The general time path contour in the complex $t$-plane with $0\leq\sigma\leq\beta$.}
\label{1}
\end{figure}

In general, thermal field theories can be defined on a general time path in the complex $t$-plane as shown in Fig. \ref{1} where $0\leq\sigma\leq\beta$ \cite{matsumoto}. In fact, the path can be generalized even further, in principle,  by adding pairs of alternating forward and backward moving real time branches, but it has been shown  \cite{nakano} that such paths are equivalent to the general path shown in Fig. \ref{1}. When $T=T'=0$, the path is associated with the imaginary time (Matsubara) formalism \cite{matsubara}. On the other hand, for real time formalisms where time takes continuous values between $-\infty\leq t\leq \infty$, one takes the limiting values $T\rightarrow -\infty, T'\rightarrow \infty$ and, for any allowed value of the parameter $\sigma$, the path leads to a real time description of the thermal field theory. When $\sigma=0$, we have the closed time path description discussed above while thermofield dynamics is associated with $\sigma=\frac{\beta}{2}$. 

For any value of $\sigma$ in the allowed range $0\leq \sigma\leq \beta$, there is a thermal field theory description and the $2\times 2$ matrix Green's function has the form in the momentum space given by \cite{matsumoto,nakano}
\begin{equation}
G^{(\sigma)} (p) = \begin{pmatrix}
G_{11}^{(\sigma)} (p) & G_{12}^{(\sigma)} (p)\\
G_{21}^{(\sigma)} (p) & G_{22}^{(\sigma)} (p)
\end{pmatrix},\label{sigmaprop0}
\end{equation}
with
\begin{align}
G^{(\sigma)}_{11} (p) & =  G_{++}(p) = G_{11}^{\rm (TFD)} (p),\nonumber \\
G^{(\sigma)}_{12} (p) & =  e^{\sigma p_{0}}G_{+-}(p) = e^{\lambda p_{0}} G_{12}^{\rm (TFD)} (p),\nonumber \\
G^{(\sigma)}_{21} (p) & =  e^{-\sigma p_{0}}G_{-+}(p) = e^{-\lambda p_{0}} G_{21}^{\rm (TFD)} (p),\nonumber \\
G^{(\sigma)}_{22} (p) & =  G_{--}(p) = G_{22}^{\rm (TFD)} (p),\label{sigmaprop}
\end{align}
where we have introduced
\begin{equation}
\lambda = \sigma - \frac{\beta}{2}.\label{lambda}
\end{equation}
We want to emphasize here that conventionally the two real branches of the path are labelled as $1,2$ for any nontrivial value of $\sigma$. Only for $\sigma =0$, namely, for the closed time path formalism, they are labelled as $\pm$. Even though for different values of $\sigma$ the Green's functions (propagators) are different, in thermal equilibrium they lead to equivalent physical results for any $\lambda$ (or $\sigma$) \cite{matsumoto,nakano} as we will also show in a simple manner in appendix {\bf A}.

Thermal field theories defined on any one parameter ($\sigma$) family of paths can be given a diagrammatic (path integral) description. However, thermofield dynamics, corresponding to $\sigma = \frac{\beta}{2}$, also has an operator description. (In fact, thermofield dynamics has its origin in an operator description as we have already pointed out.) Therefore, a natural question arises as to whether for other values of $\sigma$, we can also have an operator description of the theory in parallel to thermofield dynamics. (For example, an operator description of theories on the closed time path does not exist yet.) This has been a question of general interest since the work of Umezawa {\it et al} \cite{matsumoto}. In spite of several attempts to find an operator description, this remains an open question. In this paper, we study this question systematically and show that an operator description for any other value (other than $\sigma=\frac{\beta}{2}$) does exist indeed, with a modified inner product (different from the standard Dirac inner product) for the thermal Hilbert space. We restrict ourselves to a free massive Klein-Gordon theory in this study, but generalization to other theories is straightforward.

Our paper is organized as follows. In section \textbf{II}, we recapitulate briefly the essential ideas of thermofield dynamics and describe the well studied Bogoliubov transformation operator which takes the zero temperature (doubled) vacuum state to the thermal vacuum (vacuum state in the thermal Hilbert space). We also point out how this Bogoliubov transformation leads to the Green's function (propagator) \eqref{TFDprop} in a factorizable matrix form. In section \textbf{III}, we point out various symmetry properties of the Green's function  \eqref{sigmaprop}  for an arbitrary (allowed) value of $\sigma$ as well as its factorization which is quite useful in our attempt to construct the Bogoliubov transformation relating the thermal vacuum to the zero temperature vacuum.  In fact, in section \textbf{IV}, we use these features to construct the Bogoliubov transformation (operator) for the arbitrary parameter $0\leq\sigma\leq\beta$ which leads to the thermal Hilbert space description of the theory. We point out how the inner product of the thermal space changes for $\sigma \neq \frac{\beta}{2}$ and we show, in particular, in section \textbf{V}, how this description leads to the $2\times 2$ matrix Green's function (propagator) \eqref{sigmaprop} for the Klein-Gordon theory in a factorized manner for an  arbitrary parameter $\sigma$. We conclude with a brief summary in section \textbf{VI}. In appendix {\bf A}, we give a simple derivation of the $\lambda$ (or $\sigma$) independence of physical ensemble averages (in thermal equilibrium) in the operator formalism and point out various other features. In appendix {\bf B}, we give a brief alternative (but equivalent) operator description leading to the Green's function \eqref{sigmaprop0}-\eqref{sigmaprop}.

\section{Thermofield dynamics}
The main idea behind thermofield dynamics \cite{takahashi,umezawa,umezawa1,das} is the desire to define a thermal vacuum (and a thermal Hilbert space) so that the ensemble average of any product of operators can be written as a thermal vacuum expectation value of the operators. Namely, if there exists a state $|0(\beta)\rangle$ denoting  the thermal vacuum state, then we should be able to write
\begin{align}
\langle A_{1}\cdots A_{n}\rangle_{\beta} & = \frac{1}{Z(\beta)}\,\text{Tr}\left(e^{-\beta H} A_{1}\cdots A_{n}\right)\notag\\
& = \langle 0(\beta)|A_{1}\cdots A_{n}|0(\beta)\rangle,\label{tvac}
\end{align}
where $H$ represents the dynamical Hamiltonian for the system under study and $Z(\beta)$ stands for the partition function of the system. In this case, one can naturally develop a perturbative expansion much like at zero temperature.

With a little bit of analysis \cite{takahashi,das,khanna}, it is realized that such a state cannot be constructed if one restricts to the original Hilbert space of the theory. Rather, one needs to double the Hilbert space of the theory by adding fictitious particles known as ``tilde" particles. Let us illustrate this with the simple example of the one dimensional bosonic harmonic oscillator whose annihilation and creation operators are denoted by $a,a^{\dagger}$ and satisfy the nontrivial commutation relation
\begin{equation}
[a,a^{\dagger}] = \mathbbm{1}.
\end{equation}
For simplicity we assume that the Hamiltonian of the oscillator corresponds to the one with vanishing zero point energy so that the discussion will naturally generalize to second quantized field theories. (Having a nonvanishing zero point energy does not change the discussion.) Namely, the Hamiltonian for the system has the form
\begin{equation}
H = \omega a^{\dagger}a,
\end{equation}
where $\omega$ denotes the natural frequency of the oscillator.

Next, we double the theory by adding ``tilde" degrees of freedom through the annihilation and creation operators $\widetilde{a},\widetilde{a}^{\dagger}$ which satisfy the same commutation relations as the original oscillator degrees of freedom, namely,
\begin{equation}
[\widetilde{a}, \widetilde{a}^{\dagger}] = \mathbbm{1}.
\end{equation}
Furthermore, the two degrees of freedom are assumed to be independent so that the ``tilde" operators commute with the original operators. The Hamiltonian for the combined theory is denoted by
\begin{equation}
\widehat{H} = H - \widetilde{H} = \omega (a^{\dagger}a - \widetilde{a}^{\dagger}\widetilde{a}).\label{hhat}
\end{equation}
 
In this doubled theory, the Hilbert space is a product space of the form
\begin{equation}
|n,\widetilde{m}\rangle = |n\rangle \otimes |\widetilde{m}\rangle,
\end{equation}
where $|n\rangle, |\widetilde{n}\rangle$ denote the energy states of the two harmonic oscillator systems (namely, eigenstates of $H, \widetilde{H}$ respectively). One can construct the thermal vacuum state in this doubled space in the form
\begin{align}
|0(\beta)\rangle & = \frac{1}{Z^{\frac{1}{2}}(\beta)}\sum_{n} e^{-\frac{\beta E_{n}}{2}}|n,\widetilde{n}\rangle\notag\\
& = \frac{1}{Z^{\frac{1}{2}}(\beta)}\sum_{n} e^{-\frac{n \beta\omega}{2}}|n,\widetilde{n}\rangle,\label{thermalvac}
\end{align}
which is normalized by construction and leads to ensemble averages as thermal vacuum expectation values as desired (see \eqref{tvac}).

The thermal vacuum $|0(\beta)\rangle$ can be shown to be related to the vacuum $|0,\widetilde{0}\rangle$ of the doubled space through a Bogoliubov transformation
\begin{equation}
|0(\beta)\rangle = |0(\theta)\rangle =  U(\theta)|0,\widetilde{0}\rangle,\label{btfn}
\end{equation}
where 
\begin{equation}
U(\theta)=e^{Q(\theta)},\quad Q(\theta) = -\theta(\widetilde{a}a - a^{\dagger}\widetilde{a}^{\dagger}),\label{q}
\end{equation}
and the real parameter of the Bogoliubov transformation $\theta$ is given by
\begin{equation}
\cosh\theta = \left(e^{\beta\omega} n_{B}(\omega)\right)^{\frac{1}{2}},\quad \sinh\theta = \left(n_{B}(\omega)\right)^{\frac{1}{2}}.\label{theta}
\end{equation}
Since the generator of the Bogoliubov transformation is anti-Hermitian
\begin{equation}
Q^{\dagger}(\theta) = - Q(\theta) = Q(-\theta),\quad U^{\dagger}(\theta) = U^{-1}(\theta) = U(-\theta),\label{qprop1}
\end{equation}
it follows that the Bogoliubov transformation is unitary, namely,
\begin{equation}
U^{\dagger}(\theta) U(\theta) = \mathbbm{1} = U(\theta) U^{\dagger}(\theta).\label{qprop2}
\end{equation}
It also follows from \eqref{hhat} and \eqref{q} that
\begin{equation}
[Q(\theta), \widehat{H}] = 0,\label{qprop3}
\end{equation}
implying that the Bogoliubov transformation defines a symmetry of the doubled theory.

One can naturally define a thermal Hilbert space built on the thermal vacuum state \eqref{btfn} (or \eqref{thermalvac}). This is done by defining thermal annihilation and creation operators through the Bogoliubov transformation \eqref{q} in the following way. Let us define a doublet of operators
\begin{equation}
A = \begin{pmatrix}
a\\
\widetilde{a}^{\dagger}
\end{pmatrix}.\label{doublet}
\end{equation}
Then, with the standard commutation relations it can be derived that the Bogoliubov transformation \eqref{q} leads to the thermal doublet of operators
\begin{equation}
A(\beta) = \begin{pmatrix}
a(\beta)\\
\widetilde{a}^{\dagger}(\beta)
\end{pmatrix} = U(\theta) A U^{\dagger}(\theta) = \overline{U} (\theta) A,\label{ubar}
\end{equation}
where the $2\times 2$ matrix
\begin{equation}
\overline{U}(\theta) = \begin{pmatrix}
\cosh\theta & - \sinh\theta\\
-\sinh\theta & \cosh\theta
\end{pmatrix},\label{ubar_1}
\end{equation}
mixes up the original and the ``tilde" operators under the Bogoliubov transformation and defines the thermal operators which act on the thermal Hilbert space. We note that
\begin{equation}
\overline{U}(\theta) \sigma_{3} \overline{U}^{T}(\theta) = \sigma_{3},\label{ubar_2}
\end{equation}
where $\sigma_{3}$ denotes the third Pauli matrix. We note here that the thermal vacuum is annihilated by the thermal annihilation operators
\begin{equation}
a(\beta) |0(\beta)\rangle = 0,\quad \widetilde{a} (\beta) |0(\beta)\rangle = 0,\label{thermalvacannihilation}
\end{equation}
which leads to
\begin{equation}
\widehat{H} |0(\beta)\rangle = (H - \widetilde{H}) | 0(\beta)\rangle = 0.\label{zerothermalvacenergy}
\end{equation}
Explicitly \eqref{ubar}, \eqref{ubar_1} and \eqref{thermalvacannihilation} imply that the thermal vacuum satisfies
\begin{equation}
\cosh\theta\, a|0(\beta)\rangle - \sinh\theta\, \widetilde{a}^{\dagger} |0(\beta)\rangle = 0,\label{thermalstatecondn}
\end{equation}
which is known as the thermal state condition.

 All of these ideas from the simple example of the harmonic oscillator can be generalized to quantum field theories. The annihilation and creation operators as well as the transformation parameters simply become functions of momentum and one needs to integrate over the momentum where necessary. Thus, for the free massive Klein-Gordon theory described by the field variable $\phi (x)$, we introduce the ``tilde" field degrees of freedom described by $\widetilde{\phi} (x)$. (The ``tilde" conjugation rules \cite{ojima,das} in constructing the action for the doubled field, which we do not go into, not only replace the fields by the ``tilde" fields but also complex conjugate any coefficient.) In this case, \eqref{hhat} generalizes to
\begin{align}
\widehat{H} & = H - \widetilde{H}\notag\\
& = \int d^{3}p\ \omega_{\mathbf{p}} \left(a^{\dagger}(\mathbf{p}) a (\mathbf{p}) - \widetilde{a}^{\dagger} (\mathbf{p}) \widetilde{a} (\mathbf{p})\right),\label{hhat1}
\end{align}
where we have identified
\begin{equation}
\omega_{\mathbf{p}} = \sqrt{\mathbf{p}^{2}+m^{2}}.\label{omega}
\end{equation}

With the generator of Bogoliubov transformation (see \eqref{btfn}-\eqref{qprop3})
\begin{equation}
Q(\theta)= -\int d^{3}p\,\theta(\mathbf{p})\left(\widetilde{a}(\mathbf{p})a(\mathbf{p})-a^{\dagger}(\mathbf{p})\widetilde{a}^{\dagger}(\mathbf{p})\right),\label{q1}
\end{equation}
we can now define the thermal vacuum in the doubled space as
\begin{equation}
|0(\beta)\rangle = |0(\theta)\rangle = U(\theta)|0,\widetilde{0}\rangle = e^{Q(\theta)}|0,\widetilde{0}\rangle,\label{btfn1}
\end{equation}
where the parameters of transformation are given by
\begin{equation}
\cosh \theta(\mathbf{p}) = \left(e^{\beta\omega_{\mathbf{p}}}n_{B}(\omega_{\mathbf{p}})\right)^{\frac{1}{2}}\!\!,\ \sinh \theta(\mathbf{p}) = \left(n_{B}(\omega_{\mathbf{p}})\right)^{\frac{1}{2}}\!\!.\label{theta1}
\end{equation}
We note here that the generator of the Bogoliubov transformation is anti-Hermitian so that $U(\theta)$ is formally unitary
\begin{equation}
U^{\dagger}(\theta) U(\theta) = \mathbbm{1} = U(\theta)U^{\dagger}(\theta).\label{qprop2_1}
\end{equation} 
Furthermore,
\begin{equation}
[Q(\theta), \widehat{H}] = 0,\label{qprop3_1}
\end{equation}
as in the harmonic oscillator case so that the Bogoliubov transformation is a symmetry of the Hamiltonian of the  doubled system.

The unitary operator $U(\theta)$  allows us to construct the thermal operators in the following way. If we define a doublet of the Klein-Gordon fields 
\begin{equation}
\Phi(x) = \begin{pmatrix}
\phi(x)\\
\widetilde{\phi}(x)
\end{pmatrix},\label{Phidoublet}
\end{equation}
then with the tilde conjugation rules we can separate them into positive and negative frequency parts as
\begin{equation}
\Phi^{(\pm)} (x) = \int d^{3}p\,\frac{e^{\mp ip.x}}{\sqrt{(2\pi)^{3} 2\omega_{\mathbf{p}}}}\, \Phi^{(\pm)} (\mathbf{p}),
\end{equation}
where $p_{0}=\omega_{\mathbf{p}}$ (in the exponent) and
\begin{equation}
\Phi^{(+)} (\mathbf{p}) = \begin{pmatrix}
a (\mathbf{p})\\
\widetilde{a}^{\dagger}(\mathbf{p})
\end{pmatrix},\quad \Phi^{(-)} (\mathbf{p}) = \begin{pmatrix}
a^{\dagger}(\mathbf{p})\\
\widetilde{a}(\mathbf{p})
\end{pmatrix}.\label{Phiplusp}
\end{equation}
Under a Bogoliubov transformation, it can be checked that (see, for example, \eqref{ubar} and \eqref{ubar_1})
\begin{equation}
U(\theta) \Phi^{(\pm)} (\mathbf{p}) U^{\dagger}(\theta) = \overline{U}(\theta(\mathbf{p})) \Phi^{(\pm)} (\mathbf{p}),\label{ubar1}
\end{equation}
where the 2$\times$ 2 matrix $\overline{U}(\theta(\mathbf{p}))$ is given by
\begin{equation}
\overline{U}(\theta(\mathbf{p}))= \begin{pmatrix}
\cosh\theta(\mathbf{p}) & -\sinh\theta(\mathbf{p})\\
-\sinh\theta(\mathbf{p}) & \cosh\theta(\mathbf{p})
\end{pmatrix}.\label{ubar1_1}
\end{equation}
As in \eqref{ubar_2}, we note that
\begin{equation}
\overline{U}(\theta(\mathbf{p})) \sigma_{3}\overline{U}^{T}(\theta(\mathbf{p})) = \sigma_{3}.\label{ubar1_2}
\end{equation}
This shows that the matrices $\overline{U}(\theta)$ in \eqref{ubar_2} and $\overline{U}(\theta(\mathbf{p}))$ in \eqref{ubar1_2} belong to the group $SO(2,1)$. Furthermore, the thermal state condition \eqref{thermalstatecondn} generalizes in this case to
\begin{equation}
\cosh \theta(\mathbf{p})\, a(\mathbf{p})|0(\beta)\rangle - \sinh\theta(\mathbf{p})\,\widetilde{a}^{\dagger}(\mathbf{p})|0(\beta)\rangle = 0.\label{thermalstatecondn1}
\end{equation}

Using all these relations, the thermal $2\times 2$ matrix Green's function (propagator) of thermofield dynamics \eqref{TFDprop} can now be obtained as the expectation value in the thermal vacuum 
\begin{align}
& iG^{\rm (TFD)} (x-y) = \langle 0(\theta)|T(\Phi(x)\Phi^{T}(y))|0(\theta)\rangle \notag\\
&\  = \langle 0,\widetilde{0}|T(U (-\theta)\Phi(x)U^{\dagger}(-\theta) U(-\theta)\Phi^{T} (y) U^{\dagger}(-\theta))|0,\widetilde{0}\rangle.
\end{align}
In Fourier transformed space this leads to
\begin{equation}
G^{\rm (TFD)} (p) = \overline{U} (-\theta(\mathbf{p}))\, G^{\rm (TFD)}_{(T=0)}(p)\, \overline{U}^{T}(-\theta(\mathbf{p})),\label{TFDfactorization}
\end{equation}
where the zero temperature Green's function in thermofield dynamics has the simple form
\begin{equation}
G^{\rm (TFD)}_{(T=0)}(p) = \begin{pmatrix}
\frac{1}{p^{2}-m^{2}+i\epsilon} & 0\\
0 & - \frac{1}{p^{2}-m^{2}-i\epsilon}
\end{pmatrix},\label{zeroTprop}
\end{equation}
and $\overline{U}(\theta(\mathbf{p}))$ is the $2\times 2$ matrix defined in \eqref{ubar1_1}. Equation \eqref{TFDfactorization} is an important result. It shows that the existence of a Bogoliubov transformation leading to a thermal vacuum results in the factorization of the $2\times 2$ matrix Green's function \eqref{TFDfactorization} at finite temperature. 

\section{Factorization of propagator for an arbitrary path}

In trying to construct a thermal vacuum for an arbitrary path ($\sigma$ arbitrary), various properties of the propagator can offer helpful clues. We have already noted the form of the Green's function in \eqref{sigmaprop0} and \eqref{sigmaprop}. The components of the $2\times 2$  matrix in \eqref{sigmaprop0} 
\begin{equation}
G^{(\sigma)} (p)= \begin{pmatrix}
G^{(\sigma)}_{11}(p) & G^{(\sigma)}_{12}(p)\\
G^{(\sigma)}_{21}(p) & G^{(\sigma)}_{22}(p)
\end{pmatrix},
\end{equation}
have the explicit forms 
\begin{align}
G^{(\sigma)}_{11}(p)= & \frac{1}{p^{2}-m^{2}+i\epsilon}-2i\pi n_{B}(|p_{0}|)\delta(p^{2}-m^{2}),\nonumber \\
G^{(\sigma)}_{12}(p)= & -2i\pi e^{\lambda p_{0}}e^{\beta|p_{0}|/2}n_{B}(|p_{0}|)\delta(p^{2}-m^{2}),\nonumber \\
G^{(\sigma)}_{21}(p)= & -2i\pi e^{-\lambda p_{0}}e^{\beta|p_{0}|/2}n_{B}(|p_{0}|)\delta(p^{2}-m^{2}),\nonumber \\
G^{(\sigma)}_{22}(p)= & -\frac{1}{p^{2}-m^{2}-i\epsilon}-2i\pi n_{B}(|p_{0}|)\delta(p^{2}-m^{2}).\label{sigmaprop1}
\end{align}
and as we have noted in \eqref{lambda},  $\lambda=\sigma-\frac{\beta}{2}$.

We note from the forms of the components in \eqref{sigmaprop1} that while
\begin{equation}
G_{11}^{(\sigma)}(-p) = G_{11}^{(\sigma)} (p),\quad G_{22}^{(\sigma)}(-p) = G_{22}^{(\sigma)} (p),
\end{equation}
for any arbitrary (allowed) $\sigma$, the off-diagonal elements, in general, satisfy
\begin{equation}
G_{12}^{(\sigma)} (-p) = G_{21}^{(\sigma)} (p),\quad G_{21}^{(\sigma)} (-p) = G_{12}^{(\sigma)}(p).\label{asymmetry}
\end{equation}
Only for $\sigma=\frac{\beta}{2}$ (or $\lambda=0$), the off-diagonal matrix elements are also symmetric under $p\leftrightarrow -p$. thermofield dynamics, therefore, enjoys a very special status in that all the components of the propagator are symmetric under the reflection of the energy-momentum four vector. We will see later that this symmetry (or lack of it) is reflected in the structure of the thermal Hilbert space of the theory.

As in the case of thermofield dynamics (see \eqref{TFDfactorization} and \eqref{zeroTprop}), the matrix $G^{(\sigma)} (p)$ can also be factorized \cite{nakano} (see also \cite{evans,chu,henning,elmfors,henning1,xu,santana}), however, the factorizing matrix now depends on two parameters $(\theta(\mathbf{p}), \lambda)$. Namely, it can be checked that we can write
\begin{equation}
G^{(\sigma)}(p)= \overline{U}(-\theta(\mathbf{p}),\lambda)\,G^{\rm (TFD)}_{(T=0)} (p)\,\overline{U}^{T}(-\theta(\mathbf{p}),-\lambda),\label{sigmafactorization}
\end{equation}
where 
\begin{equation}
\overline{U}(-\theta(\mathbf{p}),\lambda)= \begin{pmatrix}
\cosh\theta(\mathbf{p}) & e^{\lambda p_{0}}\sinh\theta(\mathbf{p})\\
e^{-\lambda p_{0}}\sinh\theta(\mathbf{p}) & \cosh\theta(\mathbf{p})
\end{pmatrix}.\label{ubarthetalambda}
\end{equation}
As we have noted in the last section, the existence of a Bogoliubov transformation leading to a thermal vacuum results in the factorization of the propagator. Therefore, from \eqref{sigmafactorization} we feel that there should exist a thermal vacuum for an arbitrary $\sigma$ which can be obtained from the (doubled) zero temperature vacuum through a Bogoliubov transformation.

However, from the form of the factorizing matrix \eqref{ubarthetalambda}, we note that, while the corresponding matrix \eqref{ubar1_1} in thermofield dynamics is symmetric (under transposition), here we have 
\begin{gather}
\overline{U}(\theta(\mathbf{p}),\lambda)=\overline{U}^{T}(\theta(\mathbf{p}),-\lambda)=\overline{U}^{-1}(-\theta(\mathbf{p}),\lambda),\nonumber \\
\overline{U}(\theta(\mathbf{p}),\lambda)\overline{U}^{T}(-\theta(\mathbf{p}),-\lambda)=\mathbbm{1}.\label{eq:12}
\end{gather}
As a result, it follows that, for an arbitrary $\sigma$, the factorizing matrix satisfies
\begin{equation}
\overline{U} (\theta(\mathbf{p}),\lambda) \sigma_{3} \overline{U}^{T} (\theta(\mathbf{p}),-\lambda) = \sigma_{3},\label{ubarthetalambda_2}
\end{equation}
where $\sigma_{3}$ is the third Pauli matrix. This can be compared with \eqref{ubar1_2} and suggests that the Bogoliubov transformation taking us to the thermal vacuum, if we can determine, may have some unusual features.

The factorizing matrix $\overline{U}(-\theta(\mathbf{p}),\lambda)$ in \eqref{ubarthetalambda} can be further factorized as
\begin{equation}
\overline{U}(-\theta(\mathbf{p}),\lambda) = \overline{V}(\lambda)\overline{U}(-\theta(\mathbf{p}))\overline{V}^{-1}(\lambda),\label{furtherfactorization}
\end{equation}
where $\overline{V}(\lambda)$ can be a $2\times 2$ matrix either in the 
diagonal form $\left(\begin{array}{cc}
1 & 0\\
0 & e^{-\lambda p_{0}}
\end{array}\right)$ or in the off-diagonal form $\left(\begin{array}{cc}
0 & 1\\
e^{-\lambda p_{0}} & 0
\end{array}\right)$. We note that, in either case, we can write $\overline{V}^{-1}(\lambda) = \overline{V}^{T}(-\lambda)$. This factorization in \eqref{furtherfactorization} is very interesting because it points to the fact that the Bogoliubov transformation leading to the thermal vacuum may involve a product of operators unlike the case in thermofield dynamics.

With all this information, we are ready to construct the Bogoliubov transformation which naturally leads to the thermal vacuum of the theory and to discuss the resulting properties of the theory in the next section.

\section{Bogoliubov transformation for an arbitrary parameter}

The ensemble average of a product of operators $A_{1}\cdots A_{n}$ in thermal equilibrium is
defined as (see, for example, \eqref{tvac})
\begin{equation}
\langle A_{1}\cdots A_{n}\rangle_{\beta}= \frac{1}{Z(\beta)}\,{\rm Tr}\left(e^{-\beta H}\,A_{1}\cdots A_{n}\right),\label{ensembleaverage}
\end{equation}
where, as we have noted earlier, $\beta$ represents the inverse temperature in units of the Boltzmann constant and  $Z(\beta)$ is the partition function for the system. ``Tr'' stands for trace over a complete set of states and $H$ denotes the dynamical Hamiltonian of the system. 

To introduce an operator description for a theory defined on the one parameter family of paths, we use the cyclicity of the trace to write
\begin{equation}
{\rm Tr}\left(e^{-\beta H}A_{1}\cdots A_{n}\right) = {\rm Tr}\left(e^{-\beta H}e^{-\lambda H}A_{1}\cdots A_{n}e^{\lambda H}\right),\label{lambdatrace}
\end{equation}
where $\lambda$ is defined in \eqref{lambda}. (We note here that the cyclicity of the trace has been used earlier \cite{henning, henning1,elmfors} to introduce an arbitrary parameter into the ensemble average in a different context. The description in such a case has been called a non-Hermitian representation of thermofield dynamics.) We note from \eqref{tvac} and \eqref{lambdatrace} that the ensemble average of a product of operators can be written as the (thermal) vacuum expectation value in thermofield dynamics (TFD)
\begin{equation}
\langle A_{1}\cdots A_{n}\rangle_{\beta} = \langle 0(\beta)|e^{-\lambda H}A_{1}\cdots A_{n}e^{\lambda H}|0(\beta)\rangle.\label{tfdexpectation}
\end{equation}

At this point there are two equivalent ways of proceeding. We can either keep the doubled operators of thermofield dynamics and look for a modified thermal vacuum state that depends on the parameter $\lambda$ in addition to $\beta$ (temperature) such that the ensemble average in \eqref{tfdexpectation} can be written as an expectation value of the product of operators $A_{1}\cdots A_{n}$ in this thermal vacuum. This would be the closest in spirit to thermofield dynamics. Or, alternatively, we can keep the thermal vacuum of TFD unmodified and look for new operators to describe the doubled theory to represent the ensemble average as an expectation value of the product of new operators in the TFD vacuum. This will be closer in spirit to having two real branches of the time path with an arbitrary separation of the imaginary argument. In the main text of the paper, we will follow the first approach in detail while in appendix {\bf B} we will discuss briefly the alternative approach.

Using the definition of the thermal vacuum in TFD, $|0(\beta)\rangle$, in \eqref{btfn} (see also \eqref{btfn1}), we 
can write the ensemble average also as
\begin{align}
& \langle A_{1}\cdots A_{n}\rangle_{\beta} = 
\langle0,\widetilde{0}|U^{\dagger}(\theta)e^{-\lambda H}A_{1}\cdots A_{n}e^{\lambda H}U(\theta)|0,\widetilde{0}\rangle\nonumber \\
& = \langle0,\widetilde{0}|e^{\lambda H}U^{\dagger}(\theta)e^{-\lambda H}A_{1}\cdots A_{n}e^{\lambda H}U(\theta)e^{-\lambda H}|0,\widetilde{0}\rangle,\label{lambdatrace1}
\end{align}
where we have used the property $H|0,\widetilde{0}\rangle=0$. 

Equation \eqref{lambdatrace1} suggests that we can define a new Bogoliubov transformation operator, $U(\theta,\lambda)$, which depends on two parameters and is related to the unitary operator $U(\theta)$ of TFD by a similarity transformation as
\begin{equation}
U(\theta,\lambda)=e^{\lambda H}U(\theta)e^{-\lambda H} =  V(\lambda)U(\theta)V^{-1}(\lambda).\label{lambdabtfn}
\end{equation}
This is quite reminiscent of the factorization in \eqref{furtherfactorization} and so, in principle, one can define a thermal vacuum depending on two parameters (for the theory on the arbitrary path) through this operator $U(\theta,\lambda)$, namely,
\begin{equation}
|0(\theta,\lambda)\rangle = U (\theta,\lambda) |0, \widetilde{0}\rangle = e^{\lambda H}|0(\beta)\rangle.\label{lambdavacuum}
\end{equation}
However, there seems to be a problem with this in that the operator $U(\theta,\lambda)$ is not naively unitary (as we would expect for a Bogoliubov transformation to be), namely, since from \eqref{lambdabtfn}
\begin{equation}
U^{\dagger} (\theta,\lambda) = e^{-\lambda H} U^{\dagger}(\theta) e^{\lambda H},\label{uthetalambdadagger}
\end{equation}
it follows that
\begin{equation}
U(\theta,\lambda)U^{\dagger} (\theta,\lambda) \neq \mathbbm{1},\quad U^{\dagger} (\theta,\lambda) U(\theta,\lambda)\neq \mathbbm{1}.\label{nonunitary}
\end{equation}
As a result, the ensemble average in \eqref{lambdatrace1} cannot be written as a thermal vacuum expectation value as in thermofield dynamics (see \eqref{tvac}). However, we also note from the definition in \eqref{lambdabtfn} that
\begin{equation}
U(\theta,\lambda) U^{\dagger}(\theta,-\lambda) = \mathbbm{1} = U^{\dagger}(\theta,-\lambda) U(\theta,\lambda),\label{unitary}
\end{equation}
which is reminiscent of \eqref{ubarthetalambda_2}.

The resolution of this problem can be understood as follows and occurs in several areas of physics, most recently in the study of $PT$ symmetric theories \cite{bender,bender1} and in pseudo-Hermitian systems \cite{mostafazadeh}. Basically, properties such as Hermiticity and unitarity are defined with respect to the inner product of a Hilbert space. The conventional adjoint $A^{\dagger}$ of an operator $A$ is defined with respect to the standard Dirac inner product $\langle \cdot | \cdot \rangle$. On the other hand, with a modified inner product defined as \cite{greenwood,greenwood1}
\begin{equation}
\langle \psi | \phi\rangle_{\Lambda} = \langle \psi | \Lambda |\phi\rangle,\label{modifiedproduct}
\end{equation}
the modified adjoint $A^{\ddagger}$ is defined through the similarity transformation \cite{greenwood,greenwood1}
\begin{equation}
A^{\ddagger} = \Lambda^{-1} A^{\dagger} \Lambda.\label{modifiedadjoint}
\end{equation} 

Therefore, if we choose $\Lambda$ to correspond to the reflection operator
\begin{equation}
\lambda\xrightarrow{\Lambda} -\lambda,\quad \Lambda^{2} = \mathbbm{1},\quad \Lambda^{\dagger}=\Lambda = \Lambda^{-1},\label{Lambda}
\end{equation}
we have
\begin{equation}
\Lambda |0(\theta,\lambda)\rangle = |0(\theta,-\lambda)\rangle,\quad \langle 0(\theta,\lambda)|\Lambda = \langle 0(\theta,-\lambda)|.\label{Lambda1}
\end{equation}
Similarly from \eqref{modifiedadjoint} we note that the adjoint with respect to the modified inner product leads to
\begin{equation}
U^{\ddagger}(\theta,\lambda) = \Lambda^{-1}U^{\dagger}(\theta,\lambda) \Lambda = U^{\dagger}(\theta,-\lambda),
\end{equation}
so that (see \eqref{unitary})
\begin{equation}
U(\theta,\lambda) U^{\ddagger}(\theta,\lambda) = \mathbbm{1} = U^{\ddagger}(\theta,\lambda)U(\theta,\lambda).\label{unitary1}
\end{equation}
Namely, the Bogoliubov transformation is formally unitary, as it should be, but with respect to the modified inner product in \eqref{modifiedproduct}. 

It follows now that
\begin{align}
& \langle 0(\theta,\lambda)|0(\theta,\lambda)\rangle_{\Lambda} = \langle 0(\theta,-\lambda)|0(\theta,\lambda)\rangle\notag\\
&\quad = \langle 0,\widetilde{0}|U^{\dagger}(\theta,-\lambda)U(\theta,\lambda)|0,\widetilde{0}\rangle = 1,\label{vacnormalization}
\end{align}
so that the thermal vacuum is indeed normalized with respect to the modified product. Furthermore, the ensemble average \eqref{lambdatrace1} can indeed be written as a thermal vacuum expectation with this modified product,
\begin{equation}
 \langle A_{1}\cdots A_{n}\rangle_{\beta} = \langle 0(\theta,\lambda)|A_{1}\cdots A_{n}|0(\theta,\lambda)\rangle_{\Lambda}.\label{vacuumexpectation}
\end{equation}
This brings out a very important feature of the thermal Hilbert space. Namely, when $\lambda = \sigma-\frac{\beta}{2} \neq 0$, the thermal Hilbert space develops a natural modified inner product \eqref{modifiedproduct} different from the standard Dirac inner product. Only for $\lambda = 0$ (or $\sigma = \frac{\beta}{2}$), namely, only for thermofield dynamics does the Hilbert space coincide with the one with a standard Dirac inner product. In this sense thermofield dynamics enjoys a very special status in an operatorial description.

Since the Bogoliubov transformation $U(\theta,\lambda)$ is formally unitary with respect to the modified inner product (see \eqref{unitary1}), operators transform under such a transformation preserving their commutation relations. Thus, for example, if we consider the free massive Klein-Gordon theory described by 
\begin{equation}
H= \int d^{3}p\,\omega_{\mathbf{p}}\,a^{\dagger}(\mathbf{p})a(\mathbf{p}),\label{KGH}
\end{equation}
where $\omega_{\mathbf{p}} > 0$ is defined in \eqref{omega}. The annihilation operator $a(\mathbf{p})$ and the creation operator $a^{\ddagger}(\mathbf{p}) = a^{\dagger}(\mathbf{p})$ satisfy the commutation relation
\begin{equation}
[a(\mathbf{p}), a^{\dagger}(\mathbf{p}')] = \delta^{3} (\mathbf{p}-\mathbf{p}').
\end{equation}
Under a Bogoliubov transformation, these operators transform as
\begin{align}
a(\mathbf{p}) & \rightarrow U(\theta,\lambda) a(\mathbf{p}) U^{\ddagger}(\theta,\lambda),\notag\\
a^{\dagger}(\mathbf{p}) & \rightarrow U(\theta,\lambda)a^{\dagger}(\mathbf{p}) U^{\ddagger}(\theta,\lambda),
\end{align}
so that the commutation relation is preserved (see \eqref{unitary1})
\begin{align}
[a(\mathbf{p}), a^{\dagger}(\mathbf{p}')] & \rightarrow U(\theta,\lambda)[a(\mathbf{p}), a^{\dagger}(\mathbf{p}')] U^{\ddagger} (\theta,\lambda)\notag\\
& = \delta^{3}(\mathbf{p}-\mathbf{p}').
\end{align}

In the next section, we will use this operatorial description to derive the propagator for the Klein-Gordon theory for  an arbitrary $\sigma$ as well as its factorization discussed in detail in section {\bf III}.

\section{Propagator for the Klein-Gordon theory for an arbitrary parameter $\sigma$}

The Bogoliubov transformation \eqref{lambdabtfn} leading to the thermal vacuum and the thermal Hilbert space can be written in a closed form for the free massive Klein-Gordon theory in the following way. Let us denote
\begin{equation}
U(\theta,\lambda) = e^{\lambda H} U(\theta) e^{-\lambda H} = e^{\lambda H} e^{Q(\theta)} e^{-\lambda H} = e^{Q(\theta,\lambda)},\label{combination}
\end{equation}
where $H$ is given in \eqref{KGH}, $Q(\theta)$ is the generator of the Bogoliubov transformation for thermofield dynamics noted in \eqref{q1} and $Q(\theta,\lambda)$ denotes the generator of the Bogoliubov transformation for an arbitrary $\sigma$. The three exponents in \eqref{combination} can be combined using the Baker-Cambell-Hausdorff formula leading to
\begin{align}
& Q(\theta,\lambda)\nonumber \\
& =-\int d^{3}p\,\theta(\mathbf{p})\left(e^{-\lambda\omega_{\mathbf{p}}}\widetilde{a}(\mathbf{p})a(\mathbf{p})-e^{\lambda\omega_{\mathbf{p}}}a^{\dagger}(\mathbf{p})\widetilde{a}^{\dagger}(\mathbf{p})\right).\label{lambdagenerator}
\end{align}
It is clear that the generator $Q(\theta,\lambda)$ is manifestly anti-Hermitian with respect to the modified product, namely,
\begin{equation}
Q^{\ddagger}(\theta,\lambda) = Q^{\dagger}(\theta,-\lambda) = - Q(\theta,\lambda),
\end{equation}
so that \eqref{unitary} (or \eqref{unitary1}) follows. Furthermore, it also follows from \eqref{lambdagenerator} that (see also \eqref{qprop3_1})
\begin{equation}
[Q(\theta,\lambda), \widehat{H}] = 0,
\end{equation}
where $\widehat{H}$ is the Hamiltonian for the doubled theory given in \eqref{hhat1}.

As in \eqref{Phidoublet}-\eqref{Phiplusp} we can decompose the doublet of fields into positive and negative frequency parts as
\begin{equation}
\Phi(x) = \int d^{3}p\left(f_{p}^{(+)} (x) \Phi^{(+)} (p) + f_{p}^{(-)}(x) \Phi^{(-)} (p)\right),\label{decomposition}
\end{equation}
where the positive and negative frequency wave functions (solutions of the Klein-Gordon equation) are given by
\begin{equation}
f_{p}^{(\pm)} (x) = \frac{e^{\mp ip\cdot x}}{\sqrt{(2\pi)^{3} 2\omega_{\mathbf{p}}}},\label{+-wfn}
\end{equation}
with $p_{0}=\omega_{\mathbf{p}}$ in the exponent and 
\begin{equation}
\Phi^{(+)} (\mathbf{p}) = \begin{pmatrix}
a (\mathbf{p})\\
\widetilde{a}^{\dagger}(\mathbf{p})
\end{pmatrix},\quad \Phi^{(-)} (\mathbf{p}) = \begin{pmatrix}
a^{\dagger} (\mathbf{p})\\
\widetilde{a} (\mathbf{p})
\end{pmatrix}.
\end{equation}
It is now straightforward to check that they transform under a Bogoliubov transformation as
\begin{equation}
U(\theta,\lambda)\Phi^{(\pm)}(\mathbf{p})U^{\dagger}(\theta,-\lambda)= \widetilde{U}(\theta(\mathbf{p}),\pm\lambda)\Phi^{(\pm)} (\mathbf{p}),\label{plusminustfn}
\end{equation}
where 
\begin{align}
\widetilde{U}\left(\theta(\mathbf{p}),\lambda\right)= & \left(\begin{array}{cc}
\cosh\theta(\mathbf{p}) & -e^{\lambda\omega_{\mathbf{p}}}\sinh\theta(\mathbf{p})\\
-e^{-\lambda\omega_{\mathbf{p}}}\sinh\theta(\mathbf{p}) & \cosh\theta(\mathbf{p})
\end{array}\right).\label{widetildeu}
\end{align}
It follows from \eqref{plusminustfn} that the thermal state condition \eqref{thermalstatecondn}, in this case, has the form
\begin{equation}
\cosh\theta (\mathbf{p})\, a (\mathbf{p})|0(\theta,\lambda)\rangle - e^{\lambda\omega_{\mathbf{p}}}\sinh\theta(\mathbf{p})\, \widetilde{a}^{\dagger}(\mathbf{p})|0(\theta,\lambda)\rangle = 0.\label{thermalstatecondn_1}
\end{equation}

The $2\times 2$ matrix \eqref{widetildeu} has the same form as \eqref{ubarthetalambda}, which appears in the factorization of the propagator, except that the off-diagonal exponent has $\omega_{\mathbf{p}}>0$ instead of $p_{0}$ which can be positive as well as negative (there is no $p_{0}$ dependence, only dependence on $\mathbf{p}$ on the left hand side in \eqref{plusminustfn}). This is different from the case of thermofield dynamics (see, for example, \eqref{ubar1} and \eqref{TFDfactorization}).

To see how \eqref{ubarthetalambda} arises let us calculate the thermal $2\times 2$ matrix propagator as a thermal vacuum expectation. Let us recall that
\begin{align}
& iG^{(\sigma)}(x-y) = \langle 0(\theta,\lambda)|T(\Phi (x) \Phi^{T}(y))|0(\theta,\lambda)\rangle_{\Lambda}\notag\\
&\qquad = \langle 0,\widetilde{0}|T(U^{\dagger}(\theta,-\lambda)\Phi(x)\Phi^{T}(y)U(\theta,\lambda))|0,\widetilde{0}\rangle.
\label{prop1}
\end{align}
We recall from \eqref{lambdabtfn} that $U^{\dagger}(\theta,-\lambda) = U(-\theta,\lambda)$ as well as $U(\theta,\lambda) = U^{\dagger}(-\theta,-\lambda)$. Furthermore, introducing a factor of unity, $U^{\dagger}(-\theta,-\lambda)U(-\theta,\lambda)=\mathbbm{1}$, between $\Phi(x)$ and $\Phi^{T}(y)$, we can use the relations derived in \eqref{plusminustfn}. Finally, we can use the identities involving the positive and negative frequency wave functions, namely,
\begin{align}
& \int d^{3}p f_{p}^{(\pm)}(x)f_{p}^{(\mp)}(y)\notag\\
&\quad =\int\frac{d^{4}p}{(2\pi)^{4}}e^{-ip.(x-y)}2\pi\theta(\pm p_{0})\delta(p^{2}-m^{2}),\label{+-relation}
\end{align}
which brings in the $p_{0}$ dependence into the calculation.

With all these relations as well as the vacuum state conditions
\begin{align}
a(\mathbf{p})|0,\widetilde{0}\rangle & = 0 = \widetilde{a}(\mathbf{p})|0,\widetilde{0}\rangle,\notag\\ 
\langle 0,\widetilde{0}|a^{\dagger}(\mathbf{p}) & = 0 = \langle 0,\widetilde{0}|\widetilde{a}^{\dagger}(\mathbf{p}),
\end{align}
a straightforward calculation yields
\begin{equation}
G^{(\sigma)}(x-y)=  \int\frac{d^{4}p}{(2\pi)^{4}}e^{-ip.(x-y)}G^{(\sigma)}(p),\label{gsigma}
\end{equation}
where $G^{(\sigma)}(p)$ is the propagator (Green's function) in the momentum space which has the 
naturally factorized matrix form \eqref{sigmafactorization} (as \eqref{TFDfactorization} in thermofield dynamics)
\begin{equation}
G^{(\sigma)}(p)= \overline{U} (-\theta(\mathbf{p}),\lambda) G^{\rm (TFD)}_{(T=0)} (p) \overline{U}^{T} (-\theta(\mathbf{p}),-\lambda),\label{gsigma1}
\end{equation}
with $\overline{U}(-\theta(\mathbf{p}),\lambda)$ given in \eqref{ubarthetalambda}. Basically, the relation \eqref{+-relation} changes the matrix $\widetilde{U}$ in \eqref{widetildeu} to $\overline{U}$ in \eqref{ubarthetalambda}. We also point out here that, with the modified inner product, relation \eqref{ubarthetalambda_2} implies that the matrix $\overline{U}(\theta(\mathbf{p}),\lambda)$ belongs to the group $SO(2,1)$ just like the factorizing matrix in thermofield dynamics \eqref{ubar1_2}.

\section{Conclusion}

In this paper we have shown systematically that an operator description for a theory defined on a real time path with an arbitrary $\sigma$ does indeed exist. We have constructed the Bogoliubov transformation which connects the doubled vacuum state at zero temperature to the generalized thermal vacuum. We have pointed out that, for any value of $\sigma \neq \frac{\beta}{2}$ ($0\leq \sigma\leq \beta$), the Hilbert space develops a natural modified inner product. Only for $\sigma = \frac{\beta}{2}$ corresponding to thermofield dynamics does the Hilbert space have the standard Dirac inner product.

We have derived the $2\times 2$ matrix propagator (Green's function) for the Klein-Gordon theory directly from the  expectation value of field operators in this thermal vacuum and have shown that the factorization of the propagator (for arbitrary $\sigma$) also follows from the Bogoliubov transformation of the field operators. The factorizing matrix , $\overline{U}(-\theta(\mathbf{p}),\lambda)$, belongs to the group $SO(2,1)$, but only with the modified inner product. We have shown that the further factorization of $\overline{U}(-\theta(\mathbf{p}),\lambda)$ is intimately related to the product nature \eqref{lambdabtfn} of the Bogoliubov transformation operator. 

In appendix {\bf A} we give a simple derivation of the $\lambda$ (or $\sigma$) independence of physical ensemble averages in thermal equilibrium and also point out various other features. In appendix {\bf B} we give an alternative (but equivalent) operator description where operators are redefined but the thermal vacuum is kept as that of TFD.

\appendix

\section{$\lambda$ independence of physical ensemble averages}

In this appendix we will show that even though the ensemble average \eqref{tfdexpectation} in the operator formalism appears to have a $\lambda$ dependence, physical ensemble averages (thermal correlations of the original fields of the theory) are independent of $\lambda$. 

To see this, we note that using \eqref{zerothermalvacenergy} we can write \eqref{tfdexpectation} in two equivalent ways
\begin{align}
\langle A_{1}\cdots A_{n}\rangle_{\beta} & = \langle 0(\beta)|e^{-\lambda H}A_{1}\cdots A_{n}e^{\lambda H}|0(\beta)\rangle\notag\\
& = \langle 0(\beta)|e^{-\lambda \widetilde{H}}A_{1}\cdots A_{n}e^{\lambda \widetilde{H}}|0(\beta)\rangle.\label{equivalentrelation}
\end{align}
It follows now that if the operators $A_{1},A_{2},\cdots ,A_{n}$ belong to the original theory, then each of them would commute with $\widetilde{H}$ and using the second form of \eqref{equivalentrelation} we can write
\begin{equation}
\langle A_{1}\cdots A_{n}\rangle_{\beta} = \langle 0(\beta)|A_{1}\cdots A_{n}|0(\beta)\rangle,\label{lambdaindependence}
\end{equation}
where the $\lambda$ dependence completely cancels out. This shows that the physical thermal correlation functions are independent of the parameter $\lambda$.

The same conclusion also follows if all the operators $A_{1}, A_{2},\cdots,A_{n}$ belong to the doubled (auxiliary) space (namely, if they are all ``tilde" operators). In this case, $H$ will commute with each of them and using the first form of \eqref{equivalentrelation} we obtain \eqref{lambdaindependence}. This shows that thermal correlations involving only the ``tilde" operators are also independent of the parameter $\lambda$.

The difficulty comes if some of the operators $A_{1},\cdots , A_{n}$ belong to the original space and some to the doubled auxiliary space. In this mixed case, neither $H$ nor $\widetilde{H}$ will commute with all the operators in the product. As a result, the two $\lambda$ dependent factors in \eqref{equivalentrelation} cannot be commuted past all the operators to cancel out. Therefore, the mixed thermal correlation functions will depend on the parameter $\lambda$. We have already seen this explicitly in \eqref{sigmaprop0} and \eqref{sigmaprop} where we have noted that the diagonal elements of the Green's function are independent of $\lambda$ while the off-diagonal elements are $\lambda$ dependent.

\section{Alternative description}

In this appendix we will briefly describe an alternative but equivalent operatorial formalism where the operators of the doubled theory are redefined while the thermal vacuum state is taken to be the TFD vacuum.

Let us start by noting that
\begin{align}
& e^{-\lambda \widetilde{H}} a(\mathbf{p}) e^{\lambda\widetilde{H}} = a (\mathbf{p}),\notag\\
& e^{-\lambda \widetilde{H}} \widetilde{a}(\mathbf{p}) e^{\lambda\widetilde{H}} = e^{\lambda\omega_{\mathbf{p}}}\,\widetilde{a} (\mathbf{p}).\label{newaatilde}
\end{align}
It follows that if we identify
\begin{equation}
a_{1}(\mathbf{p}) = a (\mathbf{p}),\quad a_{2}(\mathbf{p}) = e^{\lambda\omega_{\mathbf{p}}} \widetilde{a} (\mathbf{p}),\label{a1a2}
\end{equation}
their adjoints (with respect to the modified inner product) have the forms
\begin{equation}
a_{1}^{\ddagger}(\mathbf{p}) = a^{\dagger}(\mathbf{p}),\quad a_{2}^{\ddagger}(\mathbf{p}) = e^{-\lambda\omega_{\mathbf{p}}} \widetilde{a}^{\dagger} (\mathbf{p}).\label{a1a2_1}
\end{equation}
Furthermore, they satisfy the standard commutation relations
\begin{equation}
[a_{1}(\mathbf{p}), a_{1}^{\ddagger}(\mathbf{p}')] = \delta^{3}(\mathbf{p}-\mathbf{p}') = [a_{2}(\mathbf{p}), a_{2}^{\ddagger}(\mathbf{p}')].\label{commutator_12}
\end{equation}

With the help of these, we can now define two new scalar field operators $\phi_{1}(x), \phi_{2}(x)$ in terms of the old scalar field operators $\phi(x), \widetilde{\phi}(x)$ as
\begin{align}
\phi_{1}(x) & = e^{-\lambda\widetilde{H}} \phi(x) e^{\lambda\widetilde{H}} = \phi(x),\notag\\
\phi_{2}(x) & = e^{-\lambda\widetilde{H}} \widetilde{\phi}(x) e^{\lambda\widetilde{H}},\label{phi1phi2}
\end{align}
which have the plane wave expansions
\begin{align}
\phi_{1}(x) & = \int d^{3}p\left(f_{p}^{(+)}(x) a_{1} (\mathbf{p}) + f_{p}^{(-)} (x) a_{1}^{\ddagger}(\mathbf{p})\right),\notag\\
\phi_{2} (x) & = \int d^{3}p\left(f_{p}^{(-)}(x) a_{2} (\mathbf{p}) + f_{p}^{(+)} (x) a_{2}^{\ddagger}(\mathbf{p})\right).
\end{align}
Here $f_{p}^{(\pm)}(x)$ denote the positive and negative frequency plane wave solutions defined in \eqref{+-wfn}. 

The factors $e^{\mp\lambda\widetilde{H}}$ in the ensemble average \eqref{equivalentrelation} (in the second line) can now be absorbed into a redefinition of the operators into the new operators so that the ensemble average becomes an expectation value of the redefined  operators in the standard TFD thermal vacuum. The well known generator of Bogoliubov transformation of thermofield dynamics \eqref{q1} can now be written in terms of these new operators as
\begin{equation}
Q(\theta) = - \int d^{3}p\,\theta(\mathbf{p})(e^{-\lambda\omega_{\mathbf{p}}}a_{2}(\mathbf{p})a_{1}(\mathbf{p}) - e^{\lambda\omega_{\mathbf{p}}}a_{1}^{\ddagger}(\mathbf{p})a_{2}^{\ddagger}(\mathbf{p})),\label{q_12}
\end{equation}
so that the TFD thermal vacuum 
\begin{equation}
|0(\beta)\rangle = U(\theta)|0,\widetilde{0}\rangle = e^{Q(\theta)}|0,\widetilde{0}\rangle,
\end{equation}
can now be thought of as consisting of a collection of ``$1,2$" particles.

The calculation of the propagator can be carried out now as was done in section {\bf V}. For example, let us define the doublet of fields
\begin{equation}
\Phi_{(12)}(x) = \int d^{3}p\left(f_{p}^{(+)} \Phi_{(12)}^{(+)}(\mathbf{p}) + f_{p}^{(-)}(x) \Phi_{(12)}^{(-)}(\mathbf{p})\right),
\end{equation}
where
\begin{equation}
\Phi_{(12)}^{(+)}(\mathbf{p}) = \begin{pmatrix}
a_{1}(\mathbf{p})\\
a_{2}^{\ddagger} (\mathbf{p})
\end{pmatrix},\quad \Phi_{(12)}^{(-)}(\mathbf{p}) = \begin{pmatrix}
a_{1}^{\ddagger}(\mathbf{p})\\
a_{2}(\mathbf{p})
\end{pmatrix}.
\end{equation}
Using \eqref{q_12} as well as the commutators \eqref{commutator_12}, we can now calculate
\begin{equation}
U(\theta)\Phi_{(12)}^{(\pm)}(\mathbf{p}) U^{\ddagger}(\theta) = \widetilde{U}(\theta(\mathbf{p}),\pm\lambda) \Phi_{(12)}^{(\pm)}(\mathbf{p}),\label{final}
\end{equation}
where $\widetilde{U}(\theta(\mathbf{p}),\lambda)$ is defined in \eqref{widetildeu} and \eqref{final} can be compared with \eqref{plusminustfn}. The $2\times 2$ matrix propagator now follows in a straightforward manner, as before,
\begin{align}
G^{(\sigma)}(x-y) & = -i\langle 0(\beta)|T(\Phi_{(12)}(x) \Phi_{(12)}^{T}(y))|0(\beta)\rangle_{\Lambda}\notag\\
& = \int \frac{d^{4}p}{(2\pi)^{4}}\,e^{-ip\cdot (x-y)}\,G^{(\sigma)}(p),
\end{align}
with
\begin{equation}
G^{(\sigma)} (p) = \overline{U}(-\theta(\mathbf{p}),\lambda) G^{TFD}_{(T=0)}(p) \overline{U}^{T}(-\theta(\mathbf{p}),-\lambda),
\end{equation}
which can be compared with \eqref{gsigma} and \eqref{gsigma1}.


\begin{thebibliography}{10}

\bibitem{schwinger}  J. Schwinger, {\em Brownian motion of a quantum oscillator},
J. Math. Phys. \textbf{2}, 407 (1964); J. Schwinger, {\em Proceedings of the 3rd Brandeis summer institute in theoretical physics (1960)}, ed. C. Moller {\em et al}..

\bibitem{bakshi} P. M. Bakshi and K. T. Mahanthappa,  {\em Expectation value formalism in quantum field theory. I}, J. Math. Phys. {\bf 4}, 1 (1963).

\bibitem{keldysh} L. V. Keldysh, {\em Diagram technique for non equilibrium
processes}, Sov. Phys. JETP, \textbf{20}, 1018 (1965).

\bibitem{takahashi} Y. Takahashi and H. Umezawa, {\em thermofield dynamics}, Collective Phenomena 
\textbf{2}, 55 (1975), also reprinted in Int. J. Mod. Phys. {\bf B 10}, 1755 (1996).

\bibitem{umezawa} H. Umezawa, H. Matsumoto, M. Tachiki, \textit{Thermo 
field dynamics and condensed states}, North-Holland Publishing Company,
(1982). 

\bibitem{umezawa1} H. Umezawa, \textit{Advanced field theory, Micro,
macro, and thermal Physics}, AIP press, New York (1995).

\bibitem{matsubara} T. Matsubara, {\em A new approach to quantum-statistical mechanics}, Prog. Theor. Phys. {\bf 14}, 351 (1955).

\bibitem{chou} K. C. Chou, Z. B. Su, B. L. Hao and L. Yu, {\em Equilibrium
and nonequilibrium formalisms made unified}, Phys. Rep. \textbf{118},
1 (1985).

\bibitem{landsman} N. P. Landsman and Ch. G. Van Weert, {Real and imaginary-time
field theory at finite temperature and density}, Phys. Rep. \textbf{145},
141 (1987).

\bibitem{bellac} M. L. Bellac, \textit{Thermal field theory},
Cambridge University Press, Cambridge, (1996).

\bibitem{das} A. Das, \textit{Finite temperature field theory},
World scientific, Singapore (1997).

\bibitem{khanna} F. C. Khanna, A. P. C. Malbouisson, J. M. C.
Malbouisson and A. E. Santana, \textit{Thermal Quantum Field Theory:
Algebraic Aspects and Applications ,} World Scientific, Singapore
(2009).

\bibitem{matsumoto} H. Matsumoto, Y. Nakano, H. Umezawa, F. Mancini
and M. Marinaro, {\em thermofield dynamics in interaction representation},
Prog. Theor. Phys. \textbf{70}, 599 (1983). 

\bibitem{nakano} H. Matsumoto, Y. Nakano and H. Umezawa, {\em An equivalence
class of quantum field theories at finite temperature}, J. Math. Phys.
\textbf{25}, 3076 (1984). 

\bibitem{ojima} I. Ojima, {\em Gauge fields at finite temperatures: Thermo 
field dynamics, the KMS condition and their
extension to gauge theories}, Ann. Phys. \textbf{137}, 1 (1981).

\bibitem{evans} T. S. Evans, I. Hardman, H. Umezawa and Y. Yamanaka,
{\em Heisenberg and interaction representation in thermofield dynamics},
J. Math. Phys. \textbf{33}, 370 (1992).

\bibitem{chu} H. Chu and H. Umezawa, {\em A Unified formalism of
thermal quantum field theory}, Int. J. Mod. Phys. \textbf{A9}, 2363
(1994).

\bibitem{henning} P. A. Henning and H. Umezawa, {\em Diagonalization
of propagators in thermofield dynamics for relativistic quantum fields},
Nucl. Phys. \textbf{B417}, 463 (1994).

\bibitem{elmfors} P. Elmfors and H. Umezawa, {\em Generalizations of
the thermal Bogoliubov transformation}, Physica \textbf{A202}, 557
(1994). 

\bibitem{henning1} P. A. Henning, {\em thermofield dynamics for quantum
fields with continuous spectrum}, Phys. Rep \textbf{253}, 235 (1995).

\bibitem{xu} H. H. Xu, {\em Bogoliubov Matrices in Thermal Field
Theory}, Commun. Theor. Phys. \textbf{26}, 289 (1996).

\bibitem{santana} A. E. Santana, F. C. Khanna, H. Chu and Y. C.
Chang,  {\em Thermal Lie groups, classical mechanics, and thermofield dynamics},
Ann. Phys. \textbf{249}, 481(1996).

\bibitem{bender} C. M. Bender and S. Boettcher, Phys. Rev. Lett. {\bf 80}, 5243 (1998).

\bibitem{bender1} C. M. Bender, S. Boettcher and P. Meisinger, J. Math. Phys. {\bf 40}, 2201 (1999).

\bibitem{mostafazadeh} A. Mostafazadeh, J. Math. Phys. {\bf 44}, 974 (2003).

\bibitem{greenwood} A. Das and L. Greenwood, {\em An alternative construction of the positive inner product in non-Hermitian quantum mechanics}, Phys. Lett. {\bf B678}, 504 (2009). 

\bibitem{greenwood1} A. Das and L. Greenwood, {\em An alternative construction of the positive inner product for pseudo-Hermitian Hamiltonians:Examples}, J. Math. Phys. {\bf 51}, 042103 (2010).

\end{thebibliography}
\end{document}